\documentclass[%
aip,jcp,amsmath,amssymb, reprint, twocolumn,
]{revtex4-2}
\usepackage{amsmath}
\usepackage{mathtools}
\usepackage{float}
\usepackage{amsfonts}
\usepackage{amssymb}
\usepackage{dcolumn} 
\usepackage{array}
\newcolumntype{P}[1]{>{\centering\arraybackslash}p{#1}}
\newcolumntype{M}[1]{>{\centering\arraybackslash}m{#1}}
\newcolumntype{C}[1]{>{\centering\arraybackslwash}p{#1}}
\usepackage{float}
\usepackage{psfrag}
\usepackage{tabularx}
\usepackage{stackengine}
\usepackage{amssymb}
\usepackage{mathtools}  
\usepackage{xfrac} 
\usepackage[T1]{fontenc}
\usepackage{graphicx}
\usepackage{diagbox}
\usepackage[caption=false]{subfig}
\usepackage{tikz}
\usetikzlibrary{positioning}
\usetikzlibrary{arrows}
\usetikzlibrary{trees}
\usepackage{amssymb}
\usetikzlibrary{decorations.pathmorphing}
\usetikzlibrary{decorations.markings}
\usetikzlibrary{automata,positioning}
\usepackage{braket}
\usepackage{hyperref}
\usepackage{rotating}
\usepackage{adjustbox}
\usepackage{ragged2e}
\usepackage{simplewick}
\usepackage{simpler-wick}

\usepackage[]{lineno}

\usepackage{csquotes}
\usepackage{physics,amsmath}
\usepackage{xcolor}
\usepackage{fancyhdr}
\UseRawInputEncoding
\usepackage[normalem]{ulem}

\usepackage{scalerel}

\begin{document}

\author{Sonaldeep Halder}
\affiliation{ Department of Chemistry,  \\ Indian Institute of Technology Bombay, \\ Powai, Mumbai 400076, India}

\author{Anish Dey}
\affiliation{ Department of Chemical Sciences,  \\ Indian Institute of Science Education and Research Kolkata, \\ West Bengal 741246, India}

\author{Chinmay Shrikhande}
\affiliation{ Department of Chemistry,  \\ Indian Institute of Technology Bombay, \\ Powai, Mumbai 400076, India}

\author{Rahul Maitra}
\email{rmaitra@chem.iitb.ac.in}
\affiliation{ Department of Chemistry,  \\ Indian Institute of Technology Bombay, \\ Powai, Mumbai 400076, India}
\affiliation{Centre of Excellence in Quantum Information, Computing, Science \& Technology, \\ Indian Institute of Technology Bombay, \\ Powai, Mumbai 400076, India}

\title{Machine Learning Assisted Cognitive Construction of a Shallow Depth Dynamic Ansatz for Noisy Quantum Hardware}

\begin{abstract}
    The development of various dynamic ansatz-constructing techniques has ushered in a new era, rendering the practical exploitation of Noisy Intermediate-Scale Quantum (NISQ) hardware for molecular simulations increasingly viable. However, they exhibit substantial measurement costs during their execution. This work involves the development of a novel protocol that capitalizes on regenerative machine learning methodologies and many-body perturbation theoretic measures to construct a highly expressive and shallow ansatz within the variational quantum eigensolver (VQE) framework. 
    The machine learning methodology is trained with the basis vectors of a low-rank expansion 
    of the $N-$electron Hilbert space to identify the dominant high-rank excited determinants 
    without requiring a large number of quantum measurements. These selected excited determinants are iteratively incorporated within the ansatz through their low-rank decomposition.
    The reduction in the number of quantum measurements and ansatz depth manifests in the robustness of our method towards hardware noise, as demonstrated through numerical applications. Furthermore, the proposed method is highly compatible with state-of-the-art neural error mitigation techniques. This approach significantly enhances the feasibility of quantum simulations in molecular systems, paving the way for impactful advancements in quantum computational chemistry.
\end{abstract}

\maketitle

\section{Introduction}

Quantum computing platforms provide an elegant solution to the formidable task posed by the exponential growth of the Hilbert space encountered in the realms of many-body physics and chemistry\cite{aspuru2005simulated,cao2019quantum, mcardle2020quantum}. Over the recent years, a plethora of state-of-the-art methods have been developed that aim to produce accurate energies and wavefunctions for molecular systems utilizing quantum hardware. Leading the pack are the variational algorithms \cite{peruzzo2014variational, delgado2021variational, grimsley2019adaptive,halder2022dual, halder2023corrections, mondal2023development,feniou2023overlap,zhao2023orbital, tang2021qubit, yordanov2021qubit, ostaszewski2021structure, tkachenko2021correlation, zhang2021adaptive, sim2021adaptive}, which rely on the dynamic construction and deployment of shallow depth parameterized ansatzes to generate the molecular wavefunctions. They are highly suitable for Noisy Intermediate-Scale Quantum (NISQ)\cite{preskill2018quantum} devices that suffer from limited coherence time, state preparation and measurement (SPAM) errors, and poor gate fidelity. However, most of these methods typically demand extensive pre-circuit measurements, significantly contributing to the computational overhead.
Additionally, noise from NISQ architecture can fundamentally alter the design of dynamic circuits. The selection of operators from the pool and the resulting unitary operation may deviate significantly from the optimal outcome as its construction is highly dependent on measurements (which have errors when utilizing NISQ hardware). Therefore, it is crucial to reduce the utilization of quantum resources when constructing dynamic ansatzes. In this regard, we should prioritize using approaches grounded in first principles or aided by machine learning. These methods have the potential to navigate around any challenges posed by the NISQ architecture, avoiding potential pitfalls. In this work, we have introduced a novel approach that combines unsupervised machine learning (ML) techniques with a first-principle-based strategy rooted in many-body perturbation theory. The outcome is a dynamically constructed ansatz that strikes an exceptional balance between compactness and expressiveness, all achieved without the burden of extensive pre-circuit measurements.

The use of neural network-based ML models to 
represent quantum states has been widespread in the realm of classical many-body 
theories and error mitigation protocols\cite{schuld2014quest, carleo2017solving, hornik1991approximation, le2008representational, choo2020fermionic, szabo2012modern, barrett2022autoregressive, han2019solving, hermann2020deep, pfau2020ab, kessler2021artificial, coe2018machine, hinton2006reducing, melko2019restricted, herzog2023solving}. The ability 
of these models to proactively learn the intricate interrelation between 
different basis functions that span the $N-$electron Hilbert space corresponding to
a quantum state can be leveraged to generate its low complexity representation in 
quantum computers.
The learned state, often called the neural quantum state (NQS), forms the 
backbone of neural quantum state tomography (NQST)\cite{torlai2018neural} and neural error
mitigation\cite{bennewitz2022neural}(NEM). This manuscript entails the utilization of 
a Restricted Boltzmann machine (RBM)\cite{montufar2018restricted, fischer2012introduction, freund1991unsupervised}, a powerful regenerative ML model, to construct an expanded 
wavefunction in terms of dominant many-body basis after learning the correlation 
from what may be ascribed as the "primary excitation subspace". Simple many-body 
perturbative measures intimately guide this process. The generated ansatz corresponding 
to this expanded wavefunction further involves the inclusion of a suite of two-body operators with 
an effective one hole - one particle excitation -- the so-called \textit{scatterers} -- resulting 
in an extremely low-depth yet highly expressive ansatz. Additionally, our method can be 
efficiently integrated with NEM, enhancing its efficacy for NISQ implementation.

 In section IIA, we briefly introduce the RBM's functioning and the generative process deployed to produce the dominant contributors (many-body basis) for the wavefunction expansion.  In section IIB, we set the background for the disentangled Unitary Coupled Cluster (dUCC) ansatz\cite{evangelista2019exact} and Variational Quantum Eigensolver (VQE)\cite{peruzzo2014variational}, which will be used to generate and optimize the wavefunction. In Section II C, we introduce an innovative protocol that combines RBM and second-order M\"{o}ller-Plesset perturbation theory (MP2) to craft a compact ansatz, leveraging the use of \textit{scatterer} operators.

\section{Theory}

\subsection{Restricted Boltzmann Machine (RBM): A Brief Overview}

\begin{figure}[!ht]
\centering
\includegraphics[width=\linewidth]{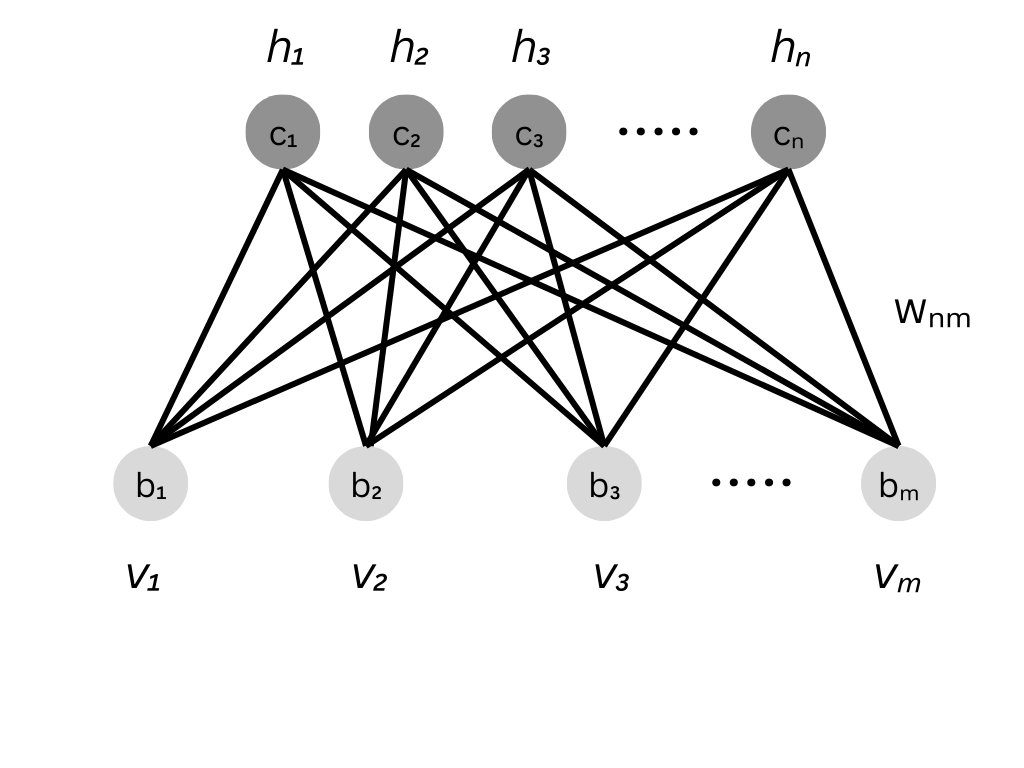}
\caption{Framework of a Restricted Boltzmann Machine with n hidden and m visible units. The biases of the visible and hidden layers are described by $\{b_i\}$ and $\{c_i\}$ respectively. $W_{nm}$ represents the weight matrix for the model. $\{v_i\}$ and $\{h_i\}$ denote the visible and hidden layer nodes respectively.}
    \label{fig:network}
\end{figure}

Characterizing a correlated wavefunction as a neural network model has been widely embraced as an effective strategy to mitigate its inherent complexity. Within this framework, the neural network's weights and biases, distributed across distinct layers, effectively encode the diverse contributions of various basis to the given wavefunction. In light of this, a regenerative neural network can effectively generate the dominant contributors to a wavefunction by training on an initial approximate state of the given system. To implement this approach, we utilize the Restricted Boltzmann Machines (RBMs). They are probabilistic graphical models that can be interpreted as stochastic neural networks. RBM comprises two layers, a visible and a hidden layer. This topology is depicted in Fig. \ref{fig:network}. The visible layer ($v_{i}\in \{0,1\}$) corresponds to observations from the training data, which comprises the binary vector representation of the basis expansion of an initial approximate wavefunction. A more detailed description of the form and source of the training dataset can be found in sections IIB and IIC. The hidden layer  ($h_{i}\in \{0,1\}$) captures the hidden patterns underlying in-between the components of the visible layer. 

To model a probability distribution of m-dimensional training data in the form of computational basis measurements, an n-dimensional hidden representation of the state is constructed. A bias is assigned to each visible and hidden unit. A weight matrix is set to establish a connection between the visible and the hidden layers, which are of dimensions $n \times m$. Connections only exist between the visible and hidden units, not between the units of the same layer. This restriction in the network topology makes RBMs different from conventional Boltzmann Machines. RBM aims to find the optimal values of the biases and the weight matrix so that the probability distribution modeled explains the observed data well. RBM constructs a joint probability distribution for the configuration $\{v,h\}$ -
\begin{equation}
    p(v,h)= \frac{1}{z} e^{-E(v,h)}
\end{equation}
where $E(v,h)$ is the Energy function and z is the partition function. 
\begin{equation}
     z= \sum_{v,h} e^{-E(v,h)} 
\end{equation}

$E(v,h)$ is parameterized by the biases and weights. The parameterized form of $E(v,h)$ is written as -

\begin{equation}
    E(v,h) = -\sum_{i=1}^{n}\sum_{j=1}^{m} w_{ij}h_{i}v_{j} -\sum_{i=1}^{n} c_{i}h_{i}
    -\sum_{j=1}^{m} b_{j}v_{j}
\end{equation}
The model is trained to minimize $E(v,h)$. To do that, the log of \textit{likelihood function} ($\mathcal{L}$) is constructed -

\begin{equation}
    \ln\mathcal{L}(\Omega|v) = \ln p(v|{\Omega}) = \ln\frac{1}{z}\sum_{h} e^{-E(v,h)}
\end{equation}
where $\Omega$ are the model's parameters. The parameters are optimized by vanishing the gradient of the log-likelihood function  
\begin{equation}
    \frac{\partial \ln\mathcal{L}(\Omega|v)}{\partial \Omega} = - \sum_{h}p(h|v) \frac{\partial E(v,h)}{\partial \Omega} +  \sum_{v,h}p(v,h) \frac{\partial E(v,h)}{\partial \Omega}
\end{equation}
where $p(h|v)$ is given as - 
\begin{equation}
    p(h|v) = \frac{e^{-E(v,h)}}{\sum_{h} e^{-E(v,h)}}
\end{equation}
which is nothing but the conditional probability of $h$ given $v$.

It must be noted that obtaining the derivative of a log-likelihood function can become intractable. Contrastive Divergence (CD), Persistent Contrastive Divergence (PCD) and Parallel Tempering (PT) are some of the approximations that are commonly employed to overcome this.

The model is trained on a collection of binary vectors (representing many-body basis), with their probability distribution ascertained by measuring an initially prepared approximate wavefunction. The hidden layers of the model decipher the correlation existing within this wavefunction. Once the training is complete, the model generates new binary vectors, preserving the correlation it learned before. The generation is accomplished through Gibbs sampling. Starting from a given binary input vector $(v_{0})$, a hidden layer representation $(h_{0})$ is calculated based on $p(h|v_{0})$. This, in turn, generates a new representation of the visible layer based on $p(v|h_{0})$. This recursive process generates new binary vectors $v^{ '}$. They correspond to the most dominant contributors to the wavefunction of a given many-fermion system. The exact parametric representation of this state is built up using the dUCC ansatz, which is further variationally optimized. This leads to section II B, where we briefly discuss the composition of dUCC ansatz and the variational principle employed to optimize the ansatz parameters.

\subsection{Disentangled Unitary Coupled Cluster Ansatz (dUCC) and Variational Quantum Eigensolver}

A trial wavefunction can be generated in a quantum computer by the action of a parameterized unitary on a reference state ($\ket{\Phi_o}$) - 

\begin{equation} \label{psi}
   \ket{\Psi (\boldsymbol{\theta})} = \hat{U}(\boldsymbol{\theta})\ket{\Phi_o}.
\end{equation}

For the dUCC ansatz, this unitary is characterized by a pool of ordered, non-commuting, anti-hermitian particle-hole operators ($\{\hat{\kappa}\}$) with the reference state taken to be the single Hartree-Fock (HF) determinant $\ket{\Phi_o} = \ket{\chi_i\chi_j...}$, $\chi$s being the spin-orbitals. The ansatz can be written as

\begin{equation}\label{6}
    \hat{U}(\boldsymbol{\theta}) = \prod_{\mu} e^{\theta_{\mu}\hat{\kappa}_{\mu}}
\end{equation}

\begin{equation}\label{kappa_break}
    \hat{\kappa}_{\mu}=\hat{\tau}_{\mu}-\hat{\tau}_{\mu}^\dagger
\end{equation}
    
\begin{equation}
    \hat{\tau}_{\mu}= \hat{a}^{\dagger}_{a}\hat{a}^{\dagger}_{b}....\hat{a}_{j}\hat{a}_{i}
\end{equation}

In the above equations, $\mu$ represents a multi-index particle-hole excitation structure as defined by the string of creation ($\hat{a}^{\dagger}$) and annihilation ($\hat{a}$) operators with the indices $\{i,j,...\}$ denoting the occupied spin-orbitals in the Hartree Fock state and $\{a,b,...\}$ denoting the unoccupied spin-orbitals. $\hat{\kappa}_{\mu}$ acts on the reference state to generate a many-body basis,

\begin{equation} \label{action_of_kappa}
    \hat{\kappa}_{\mu}\ket{\Phi_o} = \ket{\Phi_{\mu}}
\end{equation}
In a quantum computing framework, the operators described in Eq. \eqref{psi} - Eq. \eqref{action_of_kappa} are realized in terms of quantum gates and computational basis. Required transformations are carried out using standard mapping techniques. As such, these many-body bases ($\{\ket{\Phi_{\mu}}\}$) can also be represented as binary vectors (or bit strings). In this work, these vectors are often referred to as \textit{configurations}. 

For all practical applications, dUCC ansatz is constructed using only a subset of the total possible excitation operators ($\hat{\kappa}_{\mu}$). The corresponding parameters are optimized by invoking variational minimization of the electronic energy.

\begin{equation} \label{vqe_iter}
    \displaystyle{\min_{\boldsymbol{\theta}} \bra{\Phi_o}\hat{U}^{\dagger}(\boldsymbol{\theta})\hat{H}\hat{U}(\boldsymbol{\theta})\ket{\Phi_o}}
\end{equation}

In adherence to the Rayleigh-Ritz principle, the variationally obtained minimum energy gives the upper bound to the exact ground state energy ($E_o$) for the given molecular Hamiltonian. 

\begin{equation}
     \frac{\bra{\Phi_o}\hat{U}^{\dagger}(\boldsymbol{\theta})\hat{H}\hat{U}(\boldsymbol{\theta})\ket{\Phi_o}}{\bra{\Phi_o}\hat{U}^{\dagger}(\boldsymbol{\theta})\hat{U}(\boldsymbol{\theta})\ket{\Phi_o}} \ge E_o
\end{equation}

The inherent expressibility \cite{sim2019expressibility} of the chosen ansatz assumes a pivotal role in generating a trial state that recovers a substantial amount of correlation energy. It can be achieved through the incorporation of higher-order excitations (${\hat{\kappa}_{\mu}}$) in the ansatz. However, implementing such an ansatz on quantum hardware requires very deep quantum circuits. In light of the limitations posed by current quantum devices, the execution of such circuits becomes unfeasible. Developing novel techniques to generate compact and expressive ansatz becomes essential for practically utilizing quantum computing platforms for accurate molecular energy calculations.

\subsection{Utilization of RBM and Many-Body Perturbation Theory Towards the Construction of RBM-dUCC}

The construction of the dUCC ansatz, utilizing either true hole-particle or general excitations, gives rise to quantum circuits with substantial depth. Several sophisticated methodologies have been developed to incorporate dominant operators exclusively, effectively capturing a significant portion of the many-fermion correlation effects within a given molecular system. These protocols rely on quantum measurements at each stage to dynamically interlace the ansatz. This leads to a considerable measurement overhead that significantly prolongs the overall runtime of the procedure on quantum hardware. In this section, we present a comprehensive method involving judicious utilization of RBM guided by MP2 measures to achieve a compact and highly expressive ansatz. It encompasses the following steps:

\textbf{Step-1} A low-level wavefunction is constructed on a quantum device using the shallow disentangled Unitary Coupled Cluster with Singles and Doubles (dUCCSD) ansatz ($\hat{U}_{SD}(\theta)$). The parameters are optimized through variational optimization, resulting in the state $\ket{\Psi_{SD}}$. To reduce the ansatz depth, only double excitation operators with associated MP2 values above a threshold (set to be $10^{-05}$ in this work) are considered while retaining all single excitation operators.

\begin{equation}\label{optmized_SD}
    \ket{\Psi_{SD}} = \hat{U}_{SD}(\theta_{opt})\ket{\Phi_o}
\end{equation}

\begin{equation}\label{expansion}
    \ket{\Psi_{SD}} = \sum_{X\in \Sigma^m}C_{X}\ket{X}
\end{equation}

Here, $\Sigma^m$ denotes the set of bit strings of length m. Eq. \eqref{expansion} denotes the expansion of $\ket{\Psi_{SD}}$ in computational basis ($\{\ket{X}\}$). During the variational optimization of parameters associated with $\hat{U}_{SD}(\theta)$, one may choose to do a partial optimization. As will be evident in the subsequent steps, it is the relative $|C_X|^2$ (Eq. \eqref{probablility_value}) that are important and not their exact values.

\textbf{Step-2} A new ansatz, denoted as $\hat{U}_{RBM}^{dUCC}$, is derived from the probability distribution of  $\ket{\Psi_{SD}}$ (Eq. \eqref{expansion}). This ansatz organizes the excitation operators present in $\hat{U}_{SD}$ based on their associated probabilities ($\mathcal{P}_{\mu}$), sorted in descending order. Operators with corresponding probabilities below a selected threshold of $10^{-05}$ are excluded from the ansatz. The arrangement of excitation operators follows the order of rank two excitations (denoted by superscript $D$) first, followed by rank one (superscript $S$).

\begin{equation} \label{ordered ansatz}
    \hat{U}_{RBM}^{dUCC}(\boldsymbol{\theta}) = [\dots e^{\theta_2^S\hat{\kappa}_2^{S}}e^{\theta_1^S\hat{\kappa}_1^{S}}][\dots e^{\theta_2^D\hat{\kappa}_2^{D}}e^{\theta_1^D\hat{\kappa}_1^{D}}]
\end{equation}
where,
\begin{equation} \label{maps_singles}
    \{\hat{\kappa}^S\} \longrightarrow \text{Singles excitations}
\end{equation}
and
\begin{equation} \label{maps_doubles}
    \{\hat{\kappa}^D\} \longrightarrow \text{Doubles excitations}
\end{equation}
The probability order for the singles and doubles operator blocks independently follows
\begin{equation}
    \mathcal{P}_1^S > \mathcal{P}_2^S > \dots
\end{equation}
and
\begin{equation}
    \mathcal{P}_1^D > \mathcal{P}_2^D > \dots
\end{equation}
where,
\begin{equation} \label{probablility_value}
    \mathcal{P}_{\mu} = \mathcal{P}(\ket{X_{\mu}}) \propto |C_{\mu}|^2
\end{equation}
The ground and various excited determinants are mapped from the many-body basis 
to the computational basis:
\begin{equation} \label{comp_basis}
    \ket{X_{\mu}} \underset{\text{mapped from}}{\Longleftarrow} \ket{\Phi_{\mu}} 
\end{equation}
where
\begin{equation}
    \ket{\Phi_{\mu}} = \hat{\kappa}_{\mu}\ket{\Phi_o} \hspace{3mm} \forall \hat{\kappa}_{\mu} \in \hat{U}_{SD}
\end{equation}
The wavefunction generated using $\hat{U}_{RBM}^{dUCC} (\boldsymbol{\theta})$ at this stage spans what can be called as the \textit{primary excitation subspace}. An approximate 
wavefunction, which only occupies the \textit{primary excitation subspace} captures limited 
correlation.

\textbf{Step-3} RBM is trained using the computational basis bit-strings of the primary subspace ($\{\ket{X_{\mu}}\}$ in Eq. \eqref{comp_basis}) and their probabilities $\{\mathcal{P}_{\mu}\}$ incorporated in $\ket{\Psi_{SD}}$. The HF state is never taken during training since it has a high associated probability and may result in improper training\cite{herzog2023solving}. Since we are taking only a subset of the computational basis, these probabilities are further normalized. This constitutes the training phase where the ML model deciphers the correlation folded within the wavefunction. At this point, one may use powerful NEM techniques to learn a better representation of $\ket{\Psi_{SD}}$ if it has errors folded within it.

\textbf{Step-4} The trained model produces a set of new binary vectors in the computational basis. This may include vectors already present in the training set. We specifically filter out the bit-strings representing high-rank excitations ($\{\ket{Y_{\mu}}\}$) such as triples, quadruples, etc.  These high-rank excited configurations, such as triples, quadruples, etc., form what we call the \textit{secondary excitation subspace}. RBM learns the correlation that exists within the \textit{primary excitation subspace} and, accordingly, expands the wavefunction into the \textit{secondary} one through the generation of the most significant higher-order configurations. This is akin to saying that through RBM, we restrict ourselves to a very small \textit{secondary excitation subspace}, which consists of determinants that have the most dominant contribution to the molecular wavefunction.

\begin{equation}
    RBM:\{\ket{X_{\mu}}\} \mapsto \{\ket{Y_{\mu}}\}, \hspace{3mm} X,Y\in \Sigma^m
\end{equation}

$\{\ket{Y_{\mu}\}}$ represents the dominant contributors to the \textit{secondary excitation} space.

\textbf{Step-5} The obtained high-rank excitations are incorporated into $\hat{U}_{RBM}^{dUCC}$ using an indirect approach. Instead of explicitly utilizing the high-rank excitation operators that directly act on the HF state to span the secondary excitation subspace, they are induced through the action of \textit{scatterers} ($\hat{\sigma}$) on 
the primary excitation subspace functions generated previously. This implies that a given high-rank excitation is factorized into two low-rank operators. The scatterer here needs some more clarification: the inherent structural characteristics of the \textit{scatterer} enable it to act upon a set of low-rank excited determinants and generate an excitation manifold of one rank higher. Thus, the \textit{scatterers} may be perceived as two-body operators with an effective hole-particle excitation rank of one. Mathematically, these operators can be represented as:
\begin{equation}\label{def_scatterer}
\begin{split}
    \hat{\sigma}_{ki}^{cl} = a_{c}^{\dagger}a_{l}^{\dagger}a_{i}a_{k} - a_{k}^{\dagger}a_{i}^{\dagger}a_{l}a_{c}\\
    \hat{\sigma}_{ke}^{cb} = a_{c}^{\dagger}a_{b}^{\dagger}a_{e}a_{k} - a_{k}^{\dagger}a_{e}^{\dagger}a_{b}a_{c}\\
\end{split}
\end{equation}
Here,  $\{i,k,l\} \in \text{occupied orbitals}$ and $\{b,c,e\} \in \text{unoccupied orbitals}$. Unlike true excitation operators, they contain occupied to occupied or unoccupied to unoccupied \textit{transition}. This implies that the 
\textit{scatterers} have one quasi-hole or quasi-particle \textit{destruction} operator. These \textit{destruction} operators, in turn, act as a contractible set of orbitals that results in a non-vanishing commutator structure with the cluster operators with the same set of orbitals, giving rise to an effective connected 
excitation with rank one order higher than that of the cluster operator. In particular, 
if $\hat{\kappa}^D$ represents a rank two excitation operator, its commutator with a suitable \textit{scatterer} results in the
generation of a rank three excitation operator ($\hat{\kappa}^T$): 
\begin{equation}
    [\hat{\sigma}, \hat{\kappa}^D] \longrightarrow \hat{\kappa}^T
\end{equation}
The \textit{scatterers} can be configured in a nested commutator form to generate even higher-order excitations such as quadruples.
\begin{equation} \label{cascade}
\begin{split}
    [\hat{\sigma}_1, \hat{\kappa}^D] \longrightarrow \hat{\kappa}^T_{1}\\
    [\hat{\sigma}_2, \hat{\kappa}^D] \longrightarrow \hat{\kappa}^T_{2}\\
    [\hat{\sigma}_2,[\hat{\sigma}_1, \hat{\kappa}^D]] \longrightarrow \hat{\kappa}^Q
\end{split}
\end{equation}
Here, $\hat{\kappa}^D$, $\hat{\kappa}^T$, and $\hat{\kappa}^Q$ represent connected double, triple, and quadruple excitations, respectively.

Such implicit generation of the high-rank excitations is only possible when the \textit{scatterers} possess specific destruction orbital labels that are common to one of the indices of the cluster operators such that they satisfy non-commutativity. Let us say we only account for dominant triples generated by RBM. 
In that case, we take suitable non-commuting \textit{scatterers} that combine with rank two excitation operators 
already present in $\hat{U}_{RBM}^{dUCC}$ leading to the desired triples. 
The appearance of the high-rank excitations through the nested commutators is a direct consequence of the disentangled structure of the unitary (See S2 in Supplementary Material). Thus, the chosen scatterer and the
excitation operator (with which the scatterer is 
non-commutative) are paired up in a 
factorized manner.
After such incorporation, $\hat{U}_{RBM}^{dUCC}$ can be written as:
\begin{equation}
    \dots e^{\theta_2^S\hat{\kappa}_2^S}e^{\theta_1^S\hat{\kappa}_1^S} \dots (e^{\omega_3\hat{\sigma}_3}e^{\omega_2 \hat{\sigma}_2}e^{\theta_3^D \hat{\kappa}_3^D})e^{\theta_2^D\hat{\kappa}_2^D}(e^{\omega_1\hat{\sigma}_1}e^{\theta_1^D\hat{\kappa}_1^{D}})
\end{equation}
The overall depth of $\hat{U}_{RBM}^{dUCC}$ is greatly reduced due to the introduction of \textit{scatterers} and will be evident in the Results and Discussions section.

\textbf{Step-6} In step-5, the combination of distinct low rank operators with distinct \textit{scatterers} may lead to the same high rank excitation. For example, $\hat{\kappa}^D_i$ and $\hat{\kappa}^{D}_{j}$ may combine with $\hat{\sigma}_i$ and $\hat{\sigma}_{j}$ to generate the same triples excitation operator $\hat{\kappa}^T$
\begin{equation}\label{redundancy_description}
\begin{split}
    [\hat{\sigma}_i, \hat{\kappa}^D_i] \longrightarrow \hat{\kappa}^T \\
    [\hat{\sigma}_{j}, \hat{\kappa}^{D}_{j}] \longrightarrow \hat{\kappa}^T
\end{split}
\end{equation}

The most dominant combination is ascertained by the largest value of the product of the MP2 measures of the associated \textit{scatterers}, given it exceeds a predefined threshold (set to $10^{-06}$ here). The higher-order excitation ($\hat{\kappa}^T$ in Eq. \eqref{redundancy_description}) is excluded from the ansatz when no combination meets this criterion. This approach is based on the rationale that a more substantial MP2 value of the \textit{scatterer} enhances its ability to facilitate connections between the low-rank and high-rank excitations, which should be from a many-body perturbation theoretic viewpoint. Subsequently, the included high-rank excitations are reintroduced to the model (in the form of binary vectors) for further training. The probabilities for these high-rank excitations are determined by the product of two factors: the probability of the double excitation from $\ket{\Psi_{SD}}$ and the squared modulus of the MP2 values of the \textit{scatterers} involved in creating the desired high-rank excitation. Steps 4 to 6 are iterated with the improved training set until the RBM ceases to generate new high-rank configurations. This marks the termination of our protocol. The resultant ansatz ($\hat{U}_{RBM}^{dUCC}$), which we will call the RBM-dUCC ansatz, represents the final output of this protocol.
\begin{figure*}[!ht]
    \centering
\includegraphics[width=0.9\textwidth]{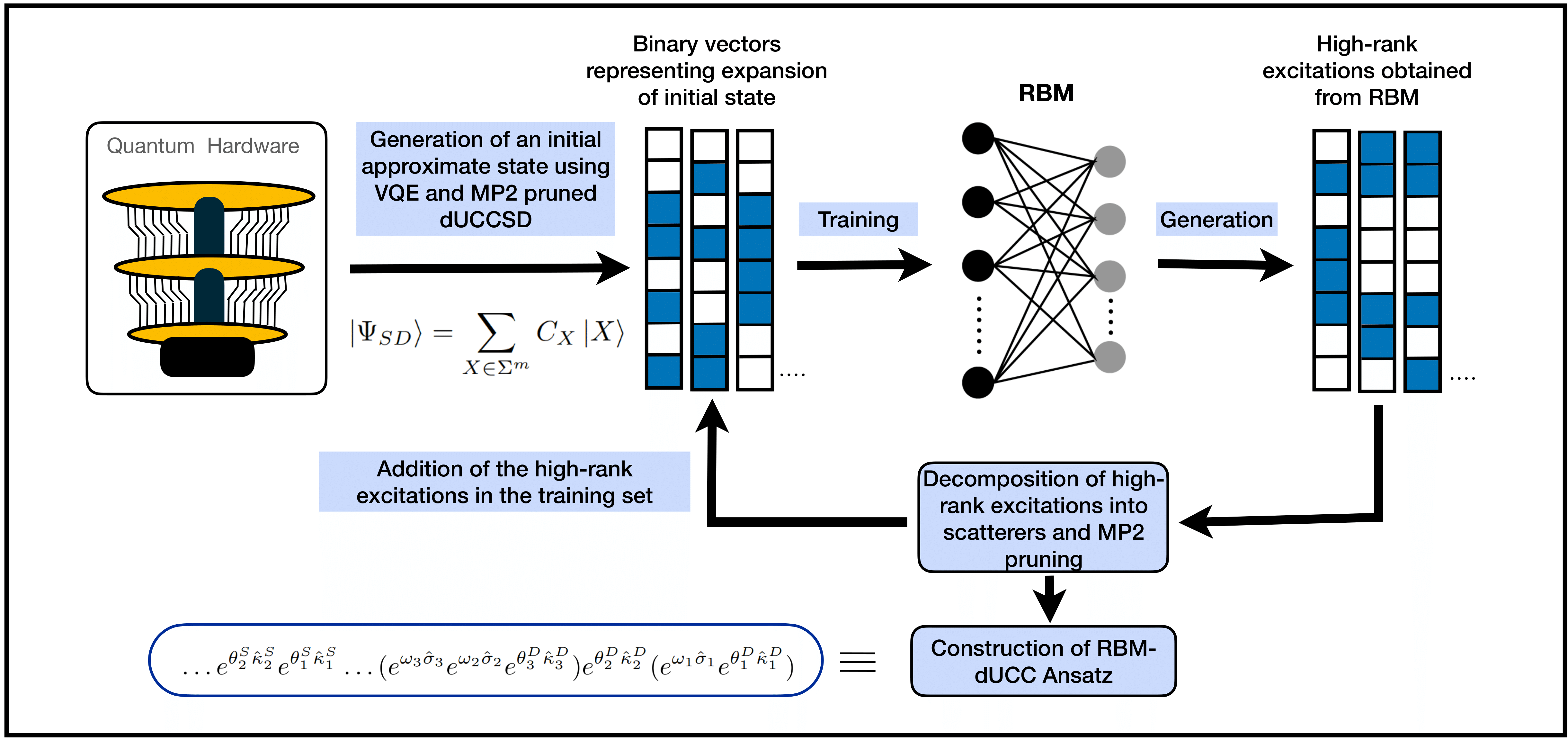}
\caption{An illustration of the protocol used to obtain shallow and highly expressive RBM-dUCC ansatz}
    \label{fig:flowchart}
\end{figure*}

The expansion of the initial wavefunction to incorporate contribution from high-order excitations occurs in steps. First, we run the sequence of the abovementioned steps, starting from $\ket{\Psi_{SD}}$ to target all dominant triply excited configurations. When RBM does not produce any new such configurations, we stop the generation procedure. The ansatz, ($\hat{U}_{RBM}^{dUCC}$),  at this stage, can be denoted as RBM-dUCCSD$T_S$, where the subscript $S$ on $T$ signifies that this ansatz generates triply excited configurations through the utilization of \textit{scatterers}. It may be noted that while we focus on generating the secondary subspace through triply excited configurations at the leading order, a \textit{scatterer} that is present later in $\hat{U}_{RBM}^{dUCC}$ may combine with it to form a quadruply excited configuration. Although we call our ansatz RBM-dUCCSD$T_S$, it may still implicitly produce some higher excited configurations, such as quadruples. This event is fortuitous and results in higher-than-desired expressibility.

The optimization of the RBM-dUCCSD$T_S$ parameters is performed in the VQE framework. The resultant state, which now expands the \textit{secondary subspace}, contains dominant triply excited configurations and may contain higher order excitations due to fortuitous combinations. This state can be fed to the protocol described in steps 2 to 6 to further produce explicit quadruples by constructing RBM-dUCCSD$T_SQ_S$. This would require explicit cascades of \textit{scatterers} as described in Eq. \eqref{cascade}. Such a procedure can be continued to make the ansatz even more expressive. It is to be noted that as the single excitation operators appear at the end, no \textit{scatterer} can act on them to produce redundant configurations.
A real quantum device has inherent noise, potentially introducing errors in obtaining $\ket{\Psi_{SD}}$.  However, once a regenerative ML model (such as RBM) learns the probability distribution, it can mitigate the errors by employing powerful NEM. As the remaining steps do not involve any quantum measurement, no additional error accumulates. The ease of integrating NEM into our method adds to its elegance.

To summarize our procedure, we first obtain a low-level approximation to the wavefunction (using a low-rank ansatz such as dUCCSD) from a quantum device and learn the probability distribution using RBM. The learned wavefunction is now expanded iteratively with the help of many body perturbative measures. We end up with a highly compact ansatz capable of inducing high-order configurations required to describe the correlation effects properly. The overview of our procedure is depicted in Fig. \ref{fig:flowchart}. In Section III, we showcase the efficacy of this method by generating dominant triply excited configurations starting from a dUCCSD ansatz. Moreover, we describe the significantly low gate depth of the constructed ansatz. We consider a variety of molecules at various geometries to carry out this study, highlighting our method's general applicability.

\section{Results and Discussions}
\subsection{Accuracy and Cost Efficiency of RBM-dUCC}
\begin{figure*}[!ht]
    \centering
\includegraphics[width=0.9\textwidth]{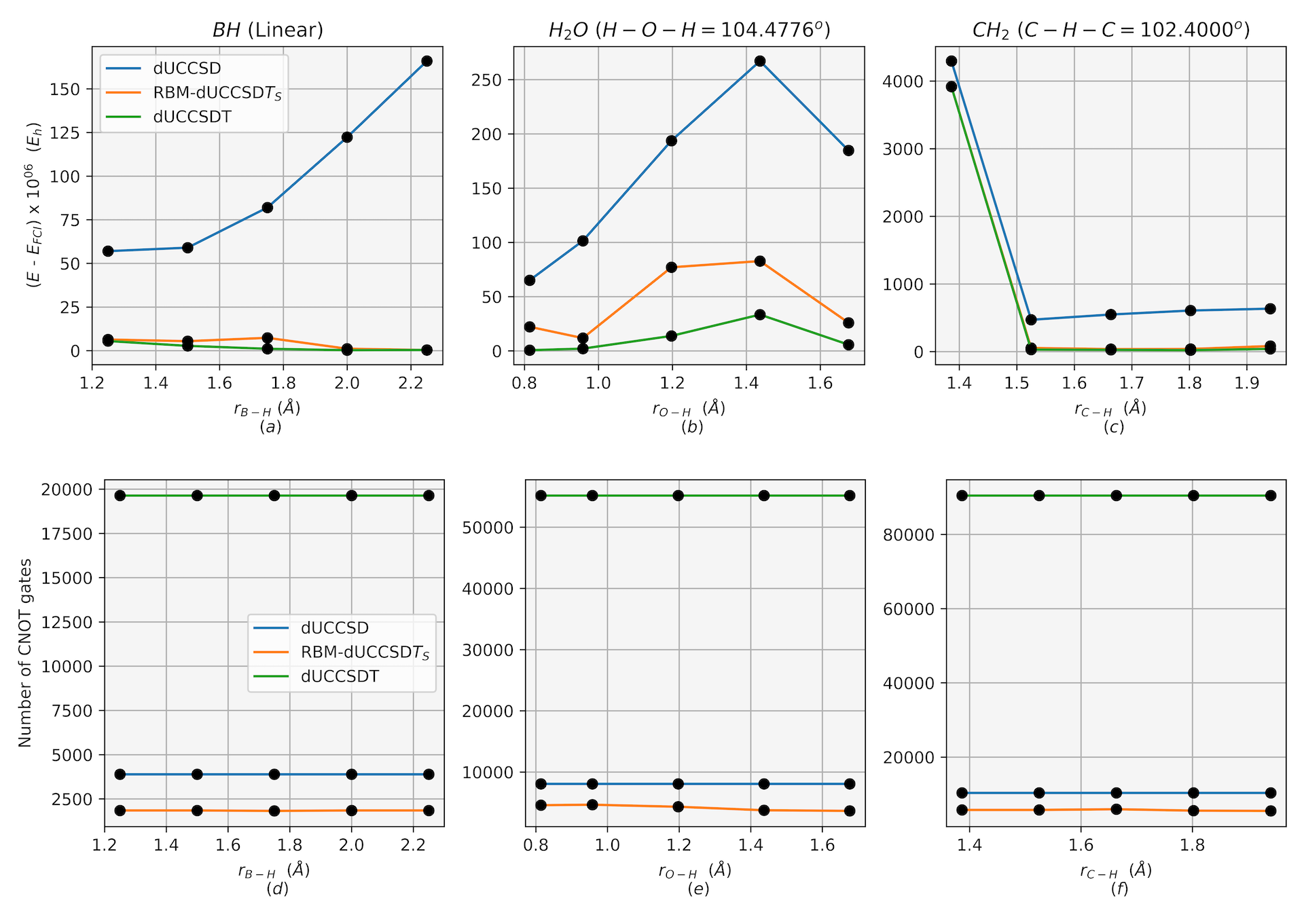}
\caption{Energy error from full configuration interaction ($(E-E_{FCI})$) for conventional dUCCSD, conventional dUCCSDT, and RBM-dUCCSD$T_S$ at different geometries of ($a$) $BH$, ($b$) $H_2O$, and ($c$) $CH_2$. Corresponding CNOT gate counts are provided in ($d$), ($e$), and ($f$), respectively.}
    \label{fig:energy_cost}
\end{figure*}

As a demonstration of the remarkable capabilities of Restricted Boltzmann Machines (RBM) in acquiring the knowledge of a wavefunction and generating dominant configurations, we compare the energy accuracy obtained using RBM-dUCCSD$T_S$ with that of conventional dUCCSDT in Fig. \ref{fig:energy_cost}. The latter consists of all triples excitations (which is of the order of $n_o^3n_v^3$, where $n_o$ represents the number of occupied and $n_v$, the number of unoccupied orbitals). We also plot the converged energies using conventional dUCCSD ansatz for reference. The comparison is depicted in Fig. \ref{fig:energy_cost} for three molecules viz. $H_2O$, $BH$ and $CH_2$ with their core orbitals frozen. For all calculations, we have used the orbitals obtained from the restricted Hartree-Fock method provided by PySCF\cite{sun2018pyscf} using the STO-3G basis set. The required Jordan-Wigner transformation for the Hamiltonian and the ansatz is obtained from the qiskit-nature\cite{Qiskit} modules. All simulations have been performed on the \textit{statevector simulator} (which mimics a noiseless quantum device) provided by qiskit. The conventional dUCCSD and dUCCSDT ansatzes and required MP2 values of the \textit{scatterers} are also obtained from qiskit modules. 
All variational optimizations have been done using a conjugate gradient (CG) optimizer with an initial point set to zero for all parameters. Of course, one may choose an initial point of zero only for the parameters associated with \textit{scatterers} and set the values of singles and doubles from the $\hat{U}_{SD}(\theta_{opt})$ (Eq. \eqref{optmized_SD}). The construction of the RBM has been carried out using the sklearn modules \cite{pedregosa2011scikit}.  A detailed description of this construction, along with its various hyper-parameters, has been included in the Supplementary Material (point S1).

\begin{figure*}[!ht]
    \centering
\includegraphics[width=0.9\textwidth]{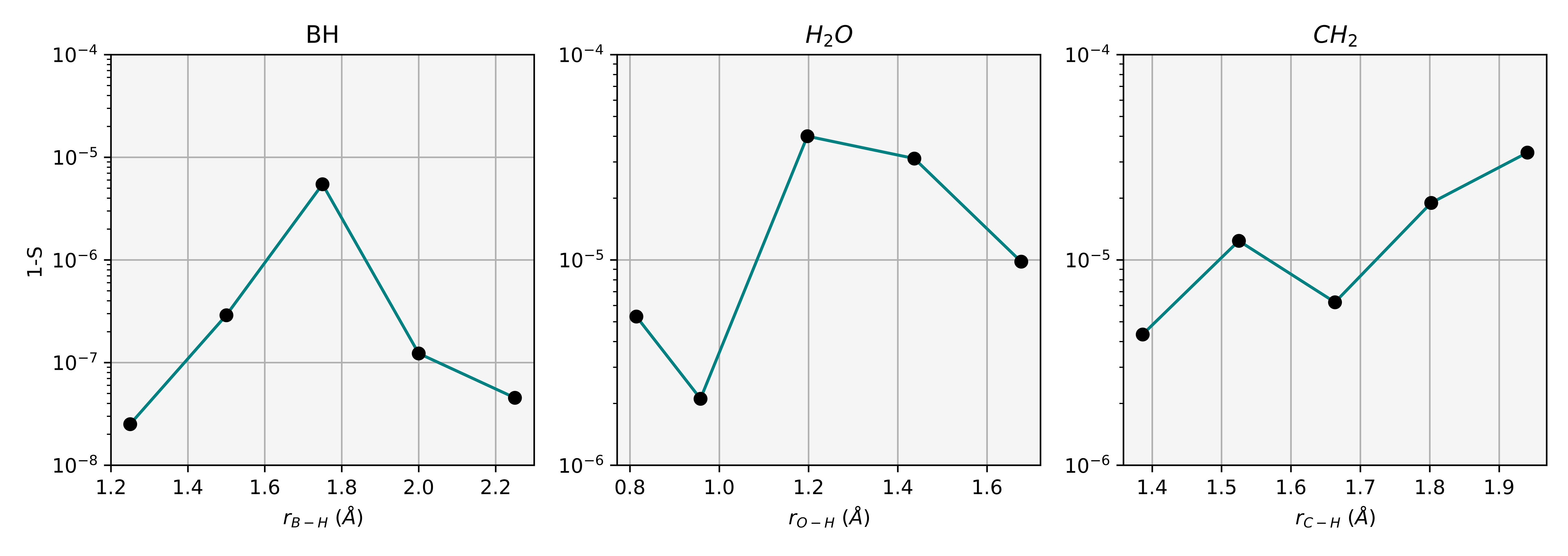}
\caption{Difference of the overlap (S) of optimized wavefunctions generated by RBM-dUCCSD$T_S$ ansatz ($\Psi_{\text{RBM}}^{\text{dUCCSD}T_S}$) and that by conventional dUCCSDT ($\Psi_{\text{Conv}}^{\text{dUCCSDT}}$) from unity. Here, S = $\braket{\Psi_{\text{RBM}}^{\text{dUCCSD}T_S}}{\Psi_{\text{Conv}}^{\text{dUCCSDT}}}$ and the molecular systems taken are identical to that used during the study depicted in Fig. \ref{fig:energy_cost}}
    \label{fig:overlap}
\end{figure*}

Apart from the energy accuracy, we also provide the circuit depth for conventional dUCCSD, dUCCSDT, and RBM-dUCCSD$T_S$. As can be discerned from Fig. \ref{fig:energy_cost}, the energy obtained after variationally optimizing RBM-dUCCSD$T_S$ is very close to that obtained using conventional dUCCSDT with a difference of $\mathcal{O}(10^{-05})$ Hartree, all while using exceptionally fewer CNOT gates (a measure of circuit depth). The former even has fewer CNOT gates than conventional dUCCSD. This tremendous reduction in the number of CNOT gates reflects the high suitability of RBM-generated ansatz for the NISQ hardware. As the energy for RBM-dUCCSD$T_S$ is not below the conventional dUCCSDT for any of the tested systems, the fortuitous generation of \textit{higher-than-desired} configurations (that is, beyond triples) has not occurred here. As we explicitly target triply excited configurations, another important assessment would be to check the overlap between the optimized wavefunctions generated using conventional dUCCSDT and RBM-dUCCSD$T_S$ ansatz. Thus, we provide the overlap between these two wavefunctions in Fig. \ref{fig:overlap}. It can be discerned from this plot that the overlap difference from unity is $\mathcal{O} (10^{-05})$ or less. This indicates the ability of RBM to recognize the dominant triply excited configurations properly. Additionally, it highlights the efficacy of MP2 theory in guiding ML predictions.

\subsection{Implementation of RBM-dUCC under Noise}

The remarkable efficiency of RBM-dUCC ansatz to produce highly correlated wavefunctions while employing only a fraction of CNOT gates has been demonstrated in Section III-A. In this section, we implement RBM-dUCCSD$T_S$ to calculate the ground state energy using VQE in a noisy backend. To construct this backend, we incorporate the shot-based \textit{estimator} \cite{Qiskit} with noise model imported from IBM's \textit{FakeMelbourne}. This includes:

\begin{enumerate}
    \item Single qubit gate errors composed of \textit{depolarizing error} channel followed by the \textit{thermal relaxation error} channel on the respective qubit.
    \item Two qubit gate errors composed of two-qubit \textit{depolarizing error} channel followed by single qubit \textit{thermal relaxation error} on the involved qubits.
    \item Single qubit readout-errors on measurement.
\end{enumerate}

\begin{figure*}[!ht]
    \centering
\includegraphics[width=0.9\textwidth]{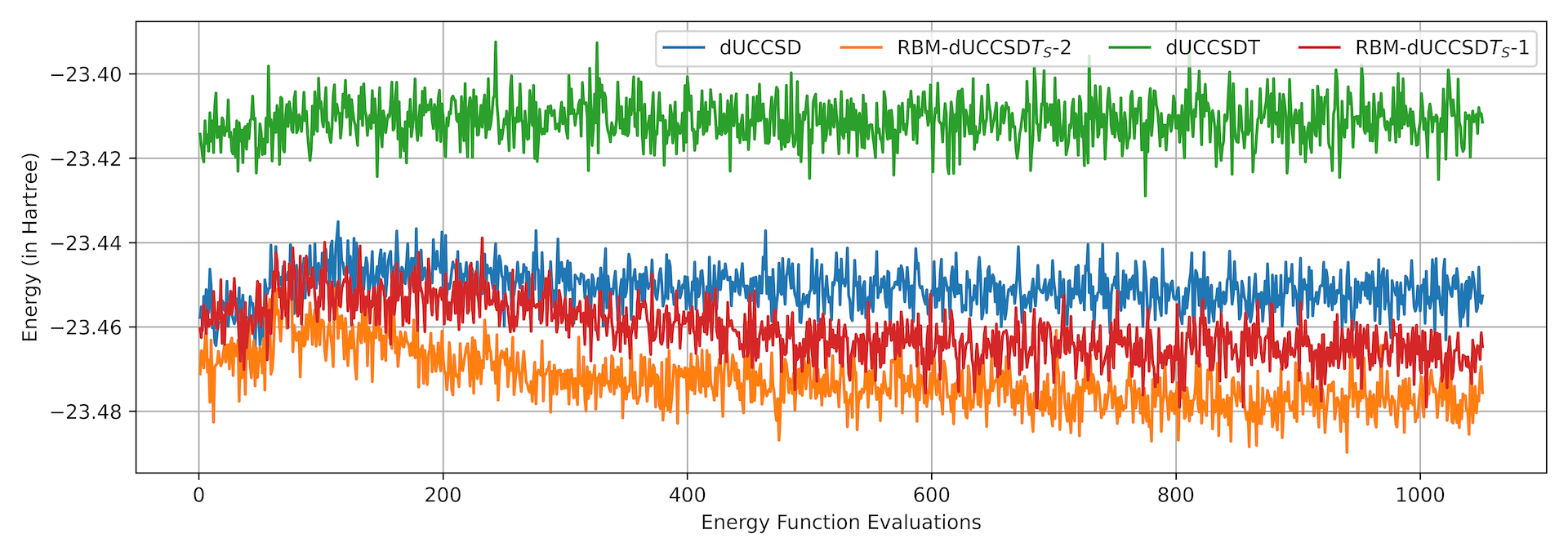}
\caption{Comparison of Energy vs. Function Evaluations for RBM-dUCCSD$T_S-1$, RBM-dUCCSD$T_S-2$, dUCCSD, and dUCCSDT for $BH$ ($r_{B-H}=2.25$\AA). The corresponding CNOT gate counts are $1912$, $1848$, $3896$, and $19640$, respectively. Each curve represents the average of data from 20 runs. For RBM-dUCCSD$T_S-1$, the initial state was optimized using an SPSA optimizer under noise conditions, as described in Section IIIB. The optimized parameters were obtained by averaging 20 runs of the VQE procedure. The initial noisy state was prepared using these parameters and fed to RBM to generate the RBM-dUCCSD$T_S-1$ ansatz.}
    \label{fig:noise_rbm}
\end{figure*}

Under the same noisy backend, we also provide a comparative performance of conventional dUCCSD and dUCCSDT ansatz. For variational optimization of the ansatz parameters, we use the Simultaneous Perturbation Stochastic Approximation (SPSA) \cite{spall1998overview} with the maximum iteration set to $500$ and the initial point taken to be zero for all parameters. In general, SPSA is considererd to be a good optimizer under noise as it requires fewer measurements during the optimization as compared to CG. The number of shots used to obtain the requisite expectation values was set to $10,000$. To reduce the time of measurement simulations, we levied the normal distribution approximation on the expectation values \cite{Qiskit}.  The resultant outcomes are portrayed through the graphical representation in Fig. \ref{fig:noise_rbm}. For this study, we used linear $BH$ ($r_{B-H}=2.25$  \AA). The large number of CNOT gates in the dUCCSDT ansatz ($\approx 19000$) results in a substantial accumulation of error. This renders the VQE optimization scheme completely useless. Even in the case of conventional dUCCSD, the trajectory is not well-behaved and leads to higher energy. RBM-dUCCSD$T_S$ provides the best results in terms of accuracy and stability. This is a direct effect of exceptionally few gates in this ansatz. We performed analysis on two variations of RBM-dUCCSD$T_S$.
\begin{enumerate}
    \item RBM-dUCCSD$T_S-1$: The initial state ($\ket{\Psi_{SD}}$) was prepared under the noise and was used to train RBM. The final ansatz obtained after the sequence of steps outlined in Section IIC was also variationally optimized under the same noise.

    \item RBM-dUCCSD$T_S-2$: The initial state ($\ket{\Psi_{SD}}$) was prepared with noiseless simulation. However, the final ansatz generated through RBM was optimized under noise.
\end{enumerate}

For both variations, the accuracy is superior to conventional dUCCSD. Although we cannot produce a noiseless initial state in real quantum hardware, our study shows the deteriorating effect of using a noisy initial wavefunction to generate the RBM-dUCC ansatz.
However, in this case, one can seamlessly use NEM to mitigate the errors and obtain a well-trained RBM. This would make the red curve (Fig. \ref{fig:noise_rbm}) move down towards the orange one. One can easily extend NEM to the final generated ansatz, too. This would further enhance the accuracy. It must be noted that in the study depicted in Fig. \ref{fig:noise_rbm}, we are not focussing on the absolute accuracy of the various ansatzes but their relative behavior on average. One must apply layers of error mitigation protocols (including NEM) to bring the results within the chemical accuracies. As the depth of the RBM-dUCC ansatz is substantially low, the cost overhead when applying such error-mitigating protocols would, in general, also decrease.

\section{Conclusions and Future Outlook}
In this study, we have demonstrated the application of Restricted Boltzmann Machines (RBMs) to acquire a concise and expressive ansatz specifically tailored to individual molecules. Our approach involves generating dominant configurations starting from an initial wavefunction derived from a lower-level approximation. To refine the output obtained from the RBM, we have employed MP2 measures as a filtering mechanism. The resultant configurations are then fed back into the RBM, enabling iterative learning and facilitating the generation of an increasingly accurate expansion of the wavefunction.
Furthermore, we have leveraged the utilization of \textit{scatterer} operators to incorporate high-rank excitations, resulting in an ansatz with exceedingly shallow depth. Notably, our methodology circumvents the necessity for quantum measurements in all steps beyond the initial wavefunction approximation. These characteristics contribute to the exceptional potency of our approach for implementation within the realm of NISQ devices. Additionally, the RBM can be combined with NEM techniques to effectively mitigate the detrimental impacts of noise inherent in present-day quantum hardware.
A meticulous examination of the various hyperparameters associated with the RBM and their influence on the final ansatz represents a promising avenue for future investigation. Furthermore, exploring alternative regenerative models beyond RBMs holds significant potential in advancing this field of study.

\section{Supplementary Material}
See the supplementary material for details about the construction of RBM and the chosen hyperparameters. It also describes the tower sampling used during the deployment of RBM. Additional information about \textit{scatterer} operators is also given.

\section{Acknowledgment}
The authors thank Dipanjali Halder (Indian Institute of Technology Bombay) for all the stimulating discussions regarding the use of \textit{scatterers} in the dUCC framework. SH acknowledges the Council of Scientific \& Industrial Research (CSIR) for their fellowship. RM acknowledges the financial support from Industrial Research and
Consultancy Centre, IIT Bombay, and
Science and Engineering Research Board, Government
of India.

\section*{AUTHOR DECLARATIONS}
\subsection*{Conflict of Interest:}
The authors have no conflict of interest to disclose.

\section*{Data Availability}
The data is available upon reasonable request to the corresponding author.

\section*{References:}

%

\end{document}